\documentstyle[manuscript,graphicx,aps,prbbib]{revtex}

\tighten
\newcommand{\be}{\begin{equation}}
\newcommand{\ee}{\end{equation}}
\newcommand{\bea}{\begin{eqnarray}}
\newcommand{\eea}{\end{eqnarray}}
\newcommand{\bd}{\begin{displaymath}}
\newcommand{\ed}{\end{displaymath}}
\newcommand{\bi}{\begin{itemize}}
\newcommand{\ei}{\end{itemize}}
\newcommand{\bc}{\begin{center}}
\newcommand{\ec}{\end{center}}
\newcommand{\bfl}{\begin{flushleft}}
\newcommand{\efl}{\end{flushleft}}
\newcommand{\bfr}{\begin{flushright}}
\newcommand{\efr}{\end{flushright}}
\newcommand{\f}{\frac}

\def\bp{{\bf p}}  \def\bq{{\bf q}} 
\def\ua{\uparrow} \def\da{\downarrow} 
\def\ra{\rightarrow} 
\def\6{\partial} \def\a{\alpha} 
 \def\d{\delta} \def\ve{\varepsilon} 
  
  \def\l{\lambda}

\def\o{\omega} \def\G{\Gamma} \def\D{\Delta}

\def\={\!\!\!&=&\!\!\!}
\def\+{\!\!\!&&\!\!\!+~}
\def\-{\!\!\!&&\!\!\!-~}

\begin{document}
\title{Field-Theoretical description of the crossover between BCS and BEC 
in a non-Fermi superconductor}
\author{C. P. Moca, I. Tifrea and M. Crisan}
\address{Department of Theoretical Physics\\
University of Cluj\\3400 Cluj, Romania}
\maketitle
\begin{abstract}
Using the field-theoretical methods we studied the evolution from BCS description 
of a non-Fermi superconductor to that of Bose-Einstein condensation (BEC) in one 
loop approximation. We showed that the repulsive interaction between composite 
bosons is determined by the exponent $\alpha$ of the Anderson propagator in a two 
dimensional model. For $\a\neq 0$ the crossover is also continous and for $\a=0$ we 
obtain the case of the Fermi liquid.
\end{abstract}

\section*{Introduction}
The problem of the crossover from BCS superconducting state to a Bose-Einstein 
condensate (BEC) of local pairs \cite{1,2,3} becomed very important in the context 
of high temperature superconductors (HTSC). While at the present time there is no 
quantitative microscopic theory for the occurrence of the superconducting state 
in the doped antiferromagnetic materials, it is generally accepted that the 
superconducting state can be described in therms of a pairing picture. The short 
coherence length ($\xi\sim$ 10-20 A) increased the interest for the problem 
\cite{4,5,6,7,8,9,10,11,12,13,14} because it showed that the BCS equations of 
highly overlapping pairs, or the description in terms of composite bosons cannot 
describe the whole regime between weak and strong coupling. The mean field method 
developed by different authors \cite{3,4,6,14} and solved analitically in two and 
three dimension, and the Ginzburg-Landau description \cite{7,8,13} showed that the 
evolution between the two limits is continous, no singularities during this evolution 
appearing. The zero temperature coherence length in the framework of 
field-theoretical method has 
been in Ref. \onlinecite{12}. The problem of the BCS-BEC crossover in 
arbitrary dimension $d$ using the field-theoretical method has been extesively 
discussed in Refs. \onlinecite{15,16,17,18}, where the chemical potential, the number 
of condensed pairs and the repulsive interaction between pairs have been calculating 
using the analogy with the field-theoretical description of superfluidity.

In this paper we apply this method to study the crossover problem for a non-Fermi 
superconductor described by the Anderson model \cite{19,20} (See also 
Refs. \onlinecite{20,21,22,23,24,25,26,27,28}), to study the crossover between weak 
coupling and strong coupling. The paper is organized as follows. In Section II we 
study the weak coupling model for a $d=2$ non-Fermi superconductor. Section III 
contains the strong coupling limit. To make the paper self-contained we present 
in Appendix the Lagrangian formalism for the superfluid phase following Refs. 
\onlinecite{15,16}. The results are discussed in Section IV.

\section*{Weak coupling limit}
The BCS-like model for the non-Fermi system is described by the Lagrangian
\be
{\cal L}=\psi^\dagger_{\ua} G_0^{-1}\psi_{\ua} + \psi^\dagger_{\da} (G_o^{-1})^*\psi_{\da} -
\l_0\psi^\dagger_{\ua}\psi^\dagger_{\da}\psi_{\da}\psi_{\ua}
\label{e1}
\ee
where the normal state is described by the Green function
\be
G_0(\bp,\o)=\f{g(\a)}{\o_c^\a(\o-\xi(\bp)+i\d)^{1-\a}}
\label{e2}
\ee
where $\o_c\leq\o\leq\o_c$, $g(\a)=\pi\a/(2\sin{(\pi\a/2)})$ and $\l_0<0$ is the coupling 
constant describing the attraction between electrons. The Green 
function given by Eq. (\ref{e2}) contain a cut-off $\o_c$ and the exponent $\a$. This form 
has been proposed first by Anderson \cite{19} to describe the 2D non-Fermi properties of the 
superconducting state.

If we introduce the two-component fermionic field
\be
\Psi=\left(\begin{array}{c} \psi_\ua\\ \psi^\dagger_\da \end{array}\right) 
\hspace{2cm}
\Psi^\dagger=\left(\begin{array}{cc} \psi^\dagger_\ua & \psi_\da \end{array}\right) 
\label{e4}
\ee
the non-interacting part of the Lagrangian (\ref{e1}) is
\be
{\cal L}_0=\Psi^\dagger\left(\begin{array}{cc}G_0^{-1} & 0 \\ 0 & (G_0^{-1})^*\end{array}\right)
\Psi
\label{e5}
\ee
In order to calculate the partition function 
\be
Z=\int D\Psi^\dagger D\Psi \exp{\left[i\int_x {\cal L}\right]}
\label{e6}
\ee
we will transform the interaction contribution from the Lagrangian (\ref{e1}) as
\be
\exp{\left[-i\l_0\int_x \psi_\ua^\dagger \psi_\da^\dagger \psi_\da\psi_\ua\right]}=
\int D\D^\dagger D\D \exp{\left[-i\int_x 
\left(\D^\dagger\psi_\da\psi_\ua+\psi_\ua^\dagger\psi_\da^\dagger\D-
\f{1}{\l_0}\D^\dagger\D\right)\right]}
\label{e7}
\ee
where $\int_x=\int dt\int d^dx$ as in Ref. \onlinecite{16,17,18}, 
and $\D=\l_0\psi_\da\psi_\ua$ is 
a bosonic field. The partition function defined by Eq. (\ref{e6}) will be expressed using Eq. 
(\ref{e7}) in a bilinear form as
\be
Z=\int D\Psi^\dagger D\Psi\int D\D^\dagger D\D \exp{\left(\f{i}{\l_0}\int_x \D^\dagger\D\right)}
\exp{\left[i\int_x\Psi^\dagger\left(\begin{array}{cc} G_0^{-1} & -\D \\
-\D^\dagger & (G_0^{-1})^*\end{array}\right)\Psi\right]}
\label{e8}
\ee
Performing the integral over the Grassmann fields the partition function becomes
\be
Z=\int D\D^\dagger D\D \exp{\left(i S_{eff}[\D^\dagger,\D]+\f{1}{\l_0}\int_x \D^\dagger \D\right)}
\label{e9}
\ee
where $S_{eff}[\D^\dagger,\D]$ is the one loop effective action, which can be written as
\be
S_{eff}[\D^\dagger,\D]=-i Tr \ln{\left(\begin{array}{cc} f(\a) (p_0-\xi(\bp))^{1-\a} & -\D\\
-\D^\dagger & f(\a) (p_0+\xi(\bp))^{1-\a}\end{array}\right)}
\label{e10}
\ee
where $f(\a)=\o_c^\a g^{-1}(\a)$ and the trace $Tr$ has been used according to the meaning 
from Ref. \onlinecite{16}.

In the mean field approximation  the integral from Eq. (\ref{e9}) can be performed using the 
solution given by the saddle point and for $T\neq 0$ the critical temperature $T_c$ will be 
obtained as
\be
T_c(\a)=\o_D\left[\f{D(\a)}{C(\a)}\right]^{1/2\a}\left[1-\f{1}{A(\a)D(\a)}\f{1}{|\l_0|}
\left(\f{\o_c}{\o_D}\right)^{2\a}\right]^{1/2\a}
\label{e11}
\ee
where $A(\a)=g^2(\a) 2^{2\a} \sin{(\pi(1-\a))}/\pi$, $C(\a)=\G^2(\a) [1-2^{1-2\a}] \zeta(\a)$, 
$D(\a)=\G(1-2\a) \G(\a)/(2\a \G(1-\a))$, $\G(x)$ being the Euler's gamma function. This 
expression is valid only in the limit $0<\a<0.5$ and a positive critical temperature 
implies for the coupling constant 
the condition $|\l_0|>\l_c$, with $\l_c=(\o_c/\o_D)^\a/(A(\a)D(\a))$.
We have to mention that the critical temperature obtained in Eq. (\ref{e11}), calculated also 
in Ref. \onlinecite{29} is different from the one obtained in Ref. \onlinecite{22,24,25}, and it is easy 
to show that it gives the exact BCS result in the limit $\a\ra 0$. If we consider the effective 
action as 
\bea
S_{eff}[\D^\dagger,\D]&=&-i Tr\ln{\left(\begin{array}{cc} f(\a)(p_0-\xi(\bp))^{1-\a} & 0\\
0 & f(\a)(p_0+\xi(\bp))^{1-\a}\end{array}\right)}\nonumber\\
&-&i Tr \ln{\left[1-\f{|\bar{\D}|^2}{f(\a)(p_0^2-\xi^2(\bp))^{1-\a}}\right]}
\label{e13}
\eea
and the system as space time independent the partition function can be written as
\be
Z=Z_0\exp{\left[\f{i}{\l_0}\bar{\D}^\dagger\bar{\D}\right]}
\label{e14}
\ee
$Z_0$ containing the non-interacting contribution, and we get for the renormalized coupling 
constant $\l$ the expression
\be
\f{1}{\l}=\f{1}{\l_0}+\f{i}{f^2(\a)}\int\f{d^2\bp}{(2\pi)^2}\int\f{dp_0}{2\pi}
\f{1}{(p_0^2-\xi^2(\bp))^{1-\a}}
\label{e15}
\ee
Using the integral
\bd
\int_{k_0}\f{1}{(k_0^2-E^2+i\eta)^l}=i(-1)^l\sqrt{\pi}\f{\G(l-1/2)}{\G(l)}\f{1}{E^{2l-1}}
\ed
we calculated $\l$ as
\be
\f{1}{\l}=\f{1}{\l_0}+\f{1}{\l_1}
\label{e17}
\ee
where
\be
\l_1=-\f{4\pi\a g^{-2}(\a)}{\cos{(\pi(\a-1))}}\f{1}{B(1/2,1/2-\a)}
\left(\f{\o_c}{\o_D}\right)^{2\a}
\label{e18}
\ee
$B(x,y)$ being the Euler beta function $B(x,y)=\G(x)\G(y)/\G(x+y)$. The expression given 
by Eq. (\ref{e18}) is positive for $\a<1/2$. The new 
coupling constant $\l$ has to be also negative in order to have superconductivity ($\l<0$) and 
this condition is satisfied if $|\l_0|<\l_1$. If we consider also the condition $\l_c<|\l_0|$ 
we get the general condition for the bare coupling constant $\l_0$, $\l_c<|\l_0|<\l_1$, 
which is satisfied for $0<\a<0.5$.

We mention that for the weak coupling limit $\l_0\ra 0^-$ the BCS limit studied in Ref. 
\onlinecite{24} is reobtained, but we also showed that the critical constant calculated from the 
critical temperature is smaller than $\l_1 (\a)$, which also satisfies condition $\l_1(\a\ra0)=0$.

In the limit $\l_0\ra -\infty$, called the strong coupling limit, we expect an important effect 
of the non-Fermi character of the electrons in the coupling constant.

\section*{Strong coupling limit}
In this limit we consider $\D(x)=\bar{\D}+\tilde{\D}(x)$ and consider the action 
$S_{eff}[\tilde{\D}^\dagger,\tilde{\D}]$ obtained from Eq. (\ref{e10}) as
\be
S_{eff}[\tilde{\D}^\dagger,\tilde{\D}]=-i Tr \ln{\left[1+\hat{G}_0\hat{\tilde{\D}}\right]}
\label{e23}
\ee
where
\be
\hat{G}_0^{-1}=\left(\begin{array}{cc} f(\a)(p_0-\xi(\bp))^{1-\a} & -\bar{\D}\\ -\bar{\D}^\dagger & 
f(\a)(p-0+\xi(\bp))^{1-\a} \end{array}\right)
\label{e24}
\ee
\be
\hat{\tilde{\D}}=
\left(\begin{array}{cc}0 & -\tilde{\D}\\ -\tilde{\D}^\dagger & 0 
\end{array}\right)
\label{e25}
\ee
which can be written as
\be
S_{eff}[\tilde{\D}^\dagger,\tilde{\D}]=-i Tr \sum_{l=1}^\infty \f{1}{l}
\left[\hat{G}_0\hat{\tilde{\D}}\right]^l
\label{e26}
\ee
with
\be
\hat{G}_0(p_0,\bp)=\f{1}{f^2(\a)(p_0^2-\xi^2(\bp))^{1-\a}-|\bar{\D}|^2}
\left(\begin{array}{cc}0 & -\tilde{\D}\\ -\tilde{\D}^\dagger & 0 
\end{array}\right)
\label{e27}
\ee
We are interested in quadratic terms in $\tilde{\D}$ and we will take the approximation
\be
S_{eff}[\tilde{\D}^\dagger,\tilde{\D}]=S_{eff}^{(2)}(0)+S_{eff}^{(2)}(\bq)
\label{e28}
\ee
which contains the quadratic contributions. The first term in Eq. (\ref{e28}) has the form 
\bea
S_{eff}^{(2)}(0)&=&\f{1}{2}i Tr \f{1}{f^2(\a)(p_0^2-\xi^2(\bp))^{1-\a}-|\bar{\D}|^2}
\left(\bar{\D}^2\tilde{\D}^\dagger\tilde{\D}^\dagger+\bar{\D}^{\dagger^2}\tilde{\D}\tilde{\D}
+2|\tilde{\D}|^2|\bar{\D}|^2\right)\\
&+&\f{1}{2}i Tr \f{1}{f^2(\a)(p_0^2-\xi^2(\bp))^{1-\a}-|\bar{\D}|^2} 2|\tilde{\D}|^2
\label{e29}
\eea
which will be approximated, taking in the dominator $\bar{\D}\approx 0$ as 
\be
S_{eff}^{(2)}(0)\cong \f{1}{2} i Tr \f{1}{f^4(\a)(p_0^2-\xi^2(\bp))^{2(1-\a)}}
\left[\bar{\D}^2\tilde{\D}^\dagger\tilde{\D}^\dagger+
\bar{\D}^{\dagger^2}\tilde{\D}\tilde{\D}+2|\tilde{\D}|^2|\bar{\D}|^2\right]
\label{e30}
\ee
the last term giving no contribution to the renormalized coupling constant. Following the 
same approximation we  calculated $S_{eff}^{(2)}(\bq)$ as 
\bea
S_{eff}^{(2)}(\bq)&=&\f{1}{2}i Tr 
\f{1}{f^2(\a)(p_0-\xi(\bp))^{1-a}(p_0+q_0+\xi(\bp+\bq))^{1-\a}}
\tilde{\D}\tilde{\D}^\dagger\nonumber\\
&+& \f{1}{2}i Tr \f{1}{f^2(\a)(p_0+\xi(\bp))^{1-\a}(p_0+q_0-\xi(\bp+\bq))^{1-\a}}
\label{e31}
\eea
From Eqs. (\ref{e30}) and (\ref{e31}) we have
\bea
{\cal L}^{(2)}(0)&=&-\f{B(1/2,3/2-2\a)}{4\pi f^4(\a)}(2m)^{3-4\a}\nonumber\\
&\times&\int\f{d^2\bp}{(2\pi)^2}\f{1}{(p^2+m\ve_a)^{3-4\a}}
\left[\bar{\D}^2\tilde{\D}^\dagger\tilde{\D}^\dagger+
\bar{\D}^{\dagger^2}\tilde{\D}\tilde{\D}+2|\tilde{\D}|^2|\bar{\D}|^2\right]
\label{e32}
\eea
and
\bea
{\cal L}^{(2)}(\bq)&=&-\f{\sin{(\pi(1-\a))B(\a,\a)}}{4\pi f^2(\a)}m^{1-2\a}
\int\f{d^2\bp}{(2\pi)^2}\f{1}{(p^2+m\ve_a+q_0m+q^2/4)^{1-2\a}}\nonumber\\
&+&-\f{\sin{(\pi(1-\a))B(\a,\a)}}{4\pi f^2(\a)}m^{1-2\a}
\int\f{d^2\bp}{(2\pi)^2}\f{1}{(p^2+m\ve_a-q_0m+q^2/4)^{1-2\a}}
\label{e33}
\eea
The integrals from Eqs. (\ref{e32}) and (\ref{e33}) can be performed using the formula
\bd
\int_\bp\f{1}{(p^2+A^2)^N}=\f{\G(N-d/2)}{(4\pi)^{d/2}\G(N)}\f{1}{(A^2)^{N-d/2}}
\ed
and we obtain
\bea
{\cal L}^{(2)}=&-&\f{m}{16\pi^2f^2(\a)}\f{2^{2-4\a}}{1-2\a}\f{B(1/2,3/2-2\a)}{\ve_a^{2-4\a}}
\left[\bar{\D}^2\tilde{\D}^\dagger\tilde{\D}^\dagger+
\bar{\D}^{\dagger^2}\tilde{\D}\tilde{\D}+2|\tilde{\D}|^2|\bar{\D}|^2\right]\nonumber\\
&+&\f{m}{16\pi^2f^2(\a)}\f{\sin{(\pi(\a-1))}}{2\a}\f{B(\a,\a)}{(\ve_a+q_0+q^2/4m)^{-2\a}}
\tilde{\D}\tilde{\D}^\dagger\nonumber\\
&+&\f{m}{16\pi^2f^2(\a)}\f{\sin{(\pi(\a-1))}}{2\a}\f{B(\a,\a)}{(\ve_a-q_0+q^2/4m)^{-2\a}}
\tilde{\D}^\dagger\tilde{\D}\nonumber\\
\label{e34}
\eea
Using the approximation
\bd
\left(\ve_a\pm q_0+\f{q^2}{4m}\right)^{2\a}\cong \ve_a^{2\a}+2\a\ve_a^{2\a-1}
\left(\pm q_0+\f{q^2}{4m}\right)
\ed
and using the notation
\be
\tilde{\Psi}=\left(\begin{array}{c} \tilde{\D} \\ \tilde{\D}^\dagger \end{array}\right)
\label{e36}
\ee
we obtain from Eq. (\ref{e34})
\be
{\cal L}^{(2)}=\f{m}{16\pi^2f^2(\a)}\sin{(\pi(1-\a))}B(\a,\a)\ve_a^{2\a-1}\f{1}{2}
\tilde{\Psi}^\dagger M \tilde{\Psi}
\label{e37}
\ee
where
\be
M=\left(\begin{array}{cc}
q_0-\f{q^2}{2m_b}-\mu_0 & -\mu_0\\ -\mu_0 & -q_0-\f{q^2}{2m_b}-\mu_0
\end{array}\right)
\label{e38}
\ee
$m_b=2m$ being the boson mass and  $\mu_0$ the chemical potential
\be
\mu_0=\f{1}{f^2(\a)}\f{2^{2-4\a}B(1/2,3/2-2\a)}{(1-2\a)\sin{(\pi(1-\a))}B(\a,\a)}
|\bar{\D}|^2\ve_a^{2\a-1}
\label{e39}
\ee
The velocity $c_0$ of the sound mode is
\be
c_0^2=\f{\mu_0}{m_b}=\f{1}{f^2(\a)}\f{2^{2-4\a}B(1/2,3/2-2\a)}{(1-2\a)\sin{(\pi(1-\a))}B(\a,\a)m}
|\bar{\D}|^2\ve_a^{2\a-1}
\label{e40}
\ee
and the repulsive interaction $\l_{0b}$ between pairs is
\be
\l_{0b}(\a)=\f{\pi^2}{m}\f{2^{4-4\a}B(1/2,3/2-2\a)}{(1-2\a)[\sin{(\pi(1-\a))}B(\a,\a)]^2}
\label{e41}
\ee
We mention that $\lim_{\a\ra 0} \l_{0b}(\a)=2\pi/m$ a result identical to the result obtained 
in Ref. \onlinecite{16} for the two dimensional case.
\section{Results and discussions}
Using the field-theoretical methods we studied the crossover between BCS and BEC in a non-Fermi 
liquid. The weak coupling case leads to the same results as in the mean field like models 
\cite{25,26,27,28}. In the strong coupling limit we showed that the pairs form a Bose gas 
with a repullsive coupling constant which is controled by $\a$.
\section{Acknowledgements}
The authors are grateful to Adriaan M. J. Schakel for the enlightening (e-mail) discussions on 
the field theoretical method.

\appendix
\section*{}
In this appendix we briefly present the field-theoretical formulation of the Bogoliubov theory 
for the interacting Bose gas. The system of an interacting Bose gas is described by the 
Lagrangean
\be
{\cal L}=\f{1}{2}\left\{\Phi^*\left[i\6_0-\hat{\ve}+\mu_0\right]\Phi-\l_0|\Phi|^4+c.c\right\}
\label{a1}
\ee
where $\Phi$ is the operator corresponding to the complex scalar field, $\hat{\ve}$ the kinetic 
energy operator, $\mu_0$ the chemical potential and $\l_0$ the bare repulsive coupling constant. 
The model described by Lagrangean (\ref{a1}) has a global symmetry to the transformation
\be
\Phi(x)\ra e^{i a} \Phi(x)
\label{a2}
\ee
where $a$ is a constant defining the transformation. At $T=0$ this symmetry is spontaneously 
broken and it is associated with the occurrence of the superfluid phase. According to the 
Goldstone theorem the dynamical restoring of the symmetry implies the occurrence of the 
Golstone bosons and it can be included in the theory taking
\be
\Phi(x)=e^{i\theta(x)}\left(\Phi_0+\tilde{\Phi}(x)\right)
\label{a3}
\ee
However, in the simple mode which is important for the pair description we will neglect the 
Goldstone mode and will take $\theta(x)=0$. The expression for the Lagrangean (\ref{a1}) 
becomes
\bea
{\cal L}=&\f{1}{2}&\left\{\tilde{\Phi}^*(x)\left[i\6_0-\hat{\ve}+\mu_0\right]\tilde{\Phi} +
\tilde{\Phi}(x)\left[-i\6_0-\hat{\ve}+\mu_0\right]\tilde{\Phi}^*(x)\right.\nonumber\\
&-&\left.\l_0 \left(\Phi^2_0\tilde{\Phi}^*(x)^2+\tilde{\Phi}^2(x)\Phi^2_0\right)-
\l_0\left(\Phi^2_0\tilde{\Phi}^2(x)+\tilde{\Phi}^*(x)^2\Phi^2_0\right)\right\}
\label{a4}
\eea
In the $p$-representation using
\be
\hat{\Phi}=\left(\begin{array}{cc}\tilde{\Phi}(x)\\ \tilde{\Phi}^*(x)\end{array}\right)
\label{a5}
\ee
Eq. (\ref{a4}) becomes
\be
{\cal L}=\f{1}{2}\hat{\Phi}^*(x)\hat{M}\hat{\Phi}(x)
\label{a6}
\ee
where
\be
\hat{M}=\left(\begin{array}{cc} p_0-\ve(\bp)-U(x)-4\l_0|\Phi_0|^2 & -2\l_0|\Phi_0|^2\\
-2\l_0|\Phi^*_0|^2 & -p_0-\ve(\bp)-U(x)-4\l_0|\Phi_0|^2\end{array}\right)
\label{a7}
\ee
and
\be
U(x)=\6_0\theta(x)+\f{1}{2m}\left[\nabla \theta(x)\right]^2
\label{a8}
\ee
Eq. (\ref{a7}) has been obtained neglecting the terms containing $\nabla^2\theta(x)$, which 
is irrelevant in the low momentum approximation, and the term ${\bf j}\nabla\theta(x)$, where 
${\bf j}$ is the current associated with $x$-variation of the phase. For the elementary 
excitations spectrum we neglect $U(x)$ (which is equivalent to neglect the $x$-dependence 
of $\theta(x)$) and from the condition
\be
det \hat{M}=0
\label{a9}
\ee
we get
\be
E^2(\bp)=\ve^2(\bp)+2\mu_0\ve(\bp)=\ve^2(\bp)+4\l_0|\Phi_0|^2\ve(\bp)
\label{a10}
\ee
The spectrum expressed by (\ref{a10}) can be approximated, in the limit $\bp\ra 0$, as
\be
E\cong u_0 |\bp|
\label{a11}
\ee
with $u_0=\sqrt{\mu_0/m}$, is gapless and it remains gapless to all orders in perturbation 
theory. For large momentum from (\ref{a10}) we get
\be
E(\bp)\cong \ve(\bp)+2\l_0|\Phi_0|^2
\label{a12}
\ee

\end{document}